\PassOptionsToPackage{table,xcdraw,prologue}{xcolor}
\documentclass[sigconf]{acmart}
\AtBeginDocument{%
  }

\usepackage{amsthm}
\usepackage{booktabs}
\usepackage{multirow}
\usepackage{algorithm}
\usepackage{algorithmic}

\copyrightyear{2025}
\acmYear{2025}
\setcopyright{acmlicensed}
\acmConference[KDD '25]{Proceedings of the 31st ACM SIGKDD Conference on Knowledge Discovery and Data Mining V.2}{August 3--7, 2025}{Toronto, ON, Canada}
\acmBooktitle{Proceedings of the 31st ACM SIGKDD Conference on Knowledge Discovery and Data Mining V.2 (KDD '25), August 3--7, 2025, Toronto, ON, Canada}
\acmDOI{10.1145/3711896.3737254}
\acmISBN{979-8-4007-1454-2/2025/08}
\begin{document}

\title{Optimizing Case-Based Reasoning System for Functional Test Script Generation with Large Language Models}


\author{Siyuan Guo}
\authornote{This work was done during the internship at Huawei.}
\affiliation{
    \department{School of Artificial Intelligence, \\ International Center of Future Science}
    \institution{Jilin University}
    \city{Changchun}
    \country{China}
}
\email{guosyjlu@gmail.com}

\author{Huiwu Liu}
\affiliation{
    \institution{Huawei Technologies Ltd.}
    \city{Nanjing}
    \country{China}
}
\email{liuhuiwu@huawei.com}

\author{Xiaolong Chen}
\affiliation{
    \institution{Huawei Technologies Ltd.}
    \city{Nanjing}
    \country{China}
}
\email{chenxiaolong42@huawei.com}

\author{Yuming Xie}
\affiliation{
    \institution{Huawei Technologies Ltd.}
    \city{Nanjing}
    \country{China}
}
\email{yuming.xie@huawei.com}

\author{Liang Zhang}
\affiliation{
    \institution{Huawei Technologies Ltd.}
    \city{Nanjing}
    \country{China}
}
\email{zhangliang1@huawei.com}

\author{Tao Han}
\affiliation{
    \institution{Huawei Technologies Ltd.}
    \city{Nanjing}
    \country{China}
}
\email{hantao@huawei.com}

\author{Hechang Chen}
\authornote{Corresponding Authors.}
\affiliation{
    \department{School of Artificial Intelligence}
    \institution{Jilin University}
    \city{Changchun}
    \country{China}
}
\email{chenhc@jlu.edu.cn}

\author{Yi Chang}
\authornotemark[2]
\affiliation{
    \department{School of Artificial Intelligence, \\ International Center of Future Science}
    \institution{Jilin University}
    \city{Changchun}
    \country{China}
}
\email{yichang@jlu.edu.cn}

\author{Jun Wang}
\authornotemark[2]
\affiliation{
    \department{AI Centre}
    \institution{University College London}
    \city{London}
    \country{UK}
}
\email{jun.wang@cs.ucl.ac.uk}

\begin{abstract}
In this work, we explore the potential of large language models (LLMs) for generating functional test scripts, which necessitates understanding the dynamically evolving code structure of the target software. 
To achieve this, we propose a case-based reasoning (CBR) system utilizing a 4R cycle (i.e., retrieve, reuse, revise, and retain), which maintains and leverages a case bank of test intent descriptions and corresponding test scripts to facilitate LLMs for test script generation.
To improve user experience further, we introduce Re4, an optimization method for the CBR system, comprising reranking-based retrieval finetuning and reinforced reuse finetuning. Specifically, we first identify positive examples with high semantic and script similarity, providing reliable pseudo-labels for finetuning the retriever model without costly labeling. Then, we apply supervised finetuning, followed by a reinforcement learning finetuning stage, to align LLMs with our production scenarios, ensuring the faithful reuse of retrieved cases.
Extensive experimental results on two product development units from Huawei Datacom demonstrate the superiority of the proposed CBR+Re4. Notably, we also show that the proposed Re4 method can help alleviate the repetitive generation issues with LLMs.
\end{abstract}

\renewcommand{\shortauthors}{Guo et al.}



\maketitle
\begin{figure}
    \includegraphics[width=0.4\textwidth]{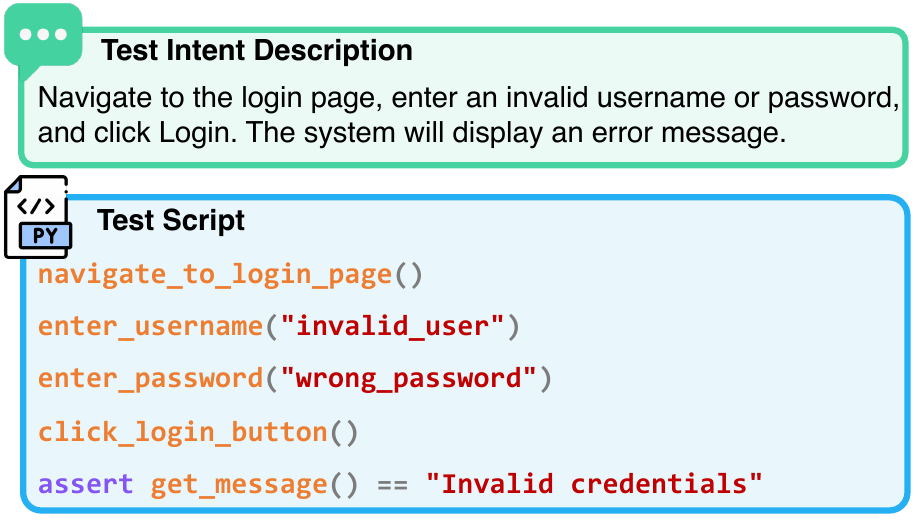}
    \caption{An example of the functional test script.}
    \label{fig:test-example}
    \vspace{-10pt}
\end{figure}

\section{Introduction}
Software testing is a critical phase in the software development lifecycle, ensuring the quality and reliability of software products. Functional testing, a key component of this process, verifies that the software’s features and operations align with the specified requirements, thereby meeting user expectations. In Huawei Datacom, writing test scripts constitutes approximately 40\% of the workload in functional testing. Even seasoned test engineers can produce only one or two test scripts per day, highlighting an urgent need to improve the efficiency of script writing.

Large language models (LLMs) have demonstrated remarkable success in complex code generation tasks, such as bug fixing \cite{swe-bench,debugbench}, automated data science \cite{DS-Agent,agentk}, and code translation \cite{unitrans,mftcoder}. Yet, their application in software testing has primarily focused on unit test generation \cite{test-eval, meta-unit-test}, leaving the more complex domain of functional testing underexplored \cite{llm4test-survey}. In this work, we investigate the potential of LLMs to assist test engineers in generating functional test scripts. Different from unit test generation, functional test scripts require invoking existing functions within the target software to build workflows based on the test intent description, as illustrated in Figure~\ref{fig:test-example}. This requirement goes beyond the static knowledge embedded in LLMs. Furthermore, as the software evolves with each version, the knowledge of the code structure should be also continuously updated. Unfortunately, such dynamic knowledge cannot be directly integrated into the context of LLMs due to the lengthy code structure of the target software, while performing continual finetuning is computationally expensive and inefficient.

\begin{figure*}
    \includegraphics[width=0.95\textwidth]{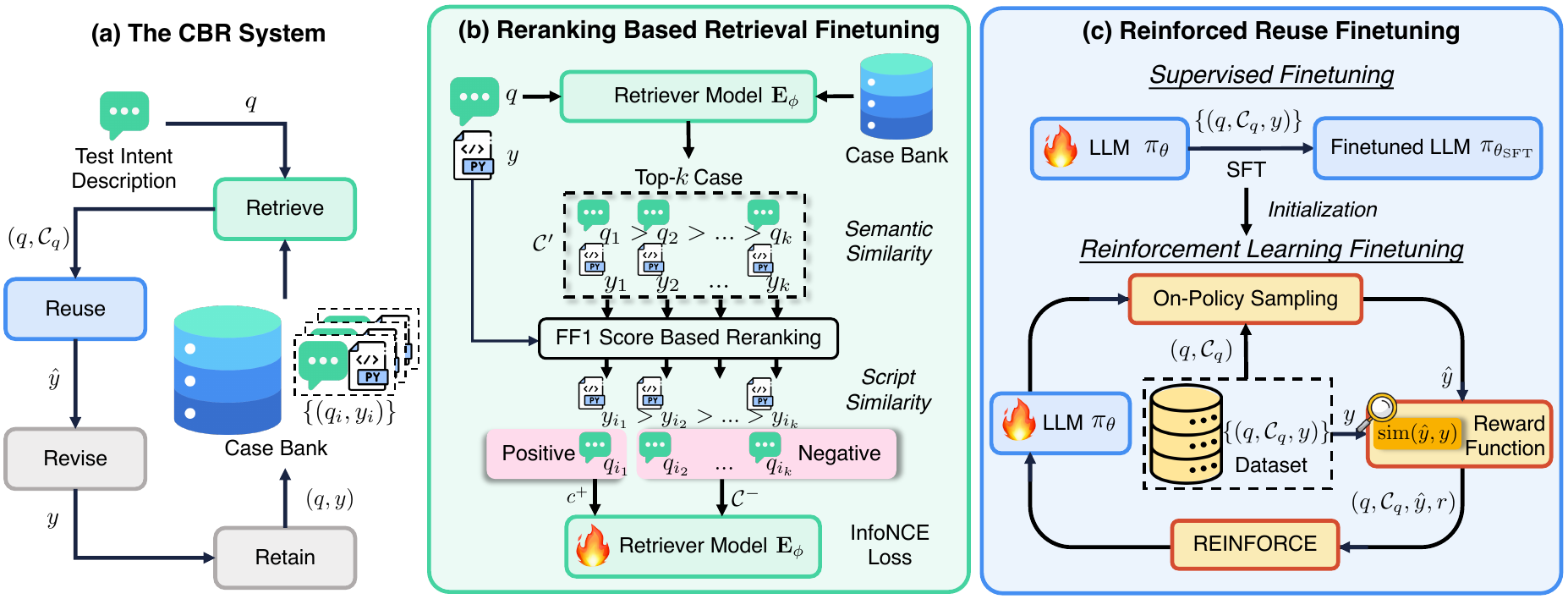}
    \caption{The overall paradigm of the proposed CBR+Re4.
    (a) The CBR system, which follows a 4R cycle of Retrieve, Reuse, Revise and Retain; (b) Reranking-based retrieval finetuning, which identifies cases with both high semantic and script similarity as positive examples and applies contrastive learning for finetuning; (c) Reinforced reuse finetuning, which consists of supervised finetuning and function f1 score based reinforcement learning finetuning for LLMs.}
    \label{fig:framework}
\end{figure*}

To address the aforementioned challenge, we adopt a classic AI problem-solving paradigm, case-based reasoning (CBR) \cite{4R-cycle, cbr-review-1, cbr-review-2, DS-Agent}, which maintains a structured case bank of past test intent descriptions and corresponding test scripts to enhance the capabilities of LLMs in test script generation. As shown in Figure~\ref{fig:framework}(a), we employ the classic 4R cycle \cite{4R-cycle} to construct the CBR system, which consists of four steps: (1) \textbf{Retrieve} similar cases from the case bank based on the given test intent description; (2) \textbf{Reuse} the retrieved cases to generate the test script using LLMs; (3) \textbf{Revise} the generated test scripts by human test engineers; (4) \textbf{Retain} the revised test script and corresponding test intent description into the case bank for future use. Benefiting from the CBR system, LLMs can utilize the mapping between test intent descriptions and corresponding function calls from the retrieved cases to generate test scripts. Additionally, the CBR system offers a flexible learning mechanism by continuously retaining human-validated cases in the case bank during the deployment stage.

Thanks to the zero-shot capabilities of pretrained retriever models and LLMs, the CBR system has shown initial effectiveness in our production scenario. To further enhance the user experience, we aim to optimize the Retrieve and Reuse steps within the CBR system, as the other steps do not rely on machine learning models. For the Retrieve step, finetuning the retriever model can be challenging due to the lack of ground-truth labeled data. While previous works propose to utilize the feedback from LLMs to generate pseudo-labels \cite{rankgpt, replug, dpa-rag}, their applicability in our production scenario is limited by the high computational and time costs associated with intensive LLM interactions. In terms of the Reuse step, supervised finetuning (SFT) appears promising for enabling LLMs to generate test scripts by reusing the retrieved cases. However, ground-truth test scripts typically include function calls absent from the retrieved cases. As a result, the SFT objective may push LLMs to generate unseen functions, introducing noise during alignment and thereby exacerbating hallucination issues during inference, which would undermine the user experience.

To this end, we propose \textbf{Re4}, an optimization method for the CBR system with \textbf{re}ranking-based \textbf{re}trieval finetuning and \textbf{re}inforced \textbf{re}use finetuning. 
Given that test script generation can be viewed as a text-to-code translation task, similar test scripts are expected to share similar test intent descriptions. Building on this insight, we propose a reranking-based retrieval finetuning method, which identifies positive examples with both high semantic and script similarity. This can provide reliable pseudo-labels for contrastive learning, enabling the finetuning of the retriever model without the need for costly human labeling or feedback from LLMs. 
Moreover, we propose a reinforced reuse finetuning method to align the LLM with our production scenario, ensuring faithful reuse of the retrieved cases for test script generation. We first perform SFT as a warm-up stage to establish initial alignment. Following this, we introduce a reinforcement learning finetuning (RLFT) stage with a critic-free online reinforcement learning algorithm, REINFORCE \cite{reinforce}. In this stage, we leverage the similarity between the generated script and the ground-truth test script as the golden reward to further refine the alignment, which eliminates the undesired behavior patterns introduced during the SFT stage. 
We present extensive experimental results to demonstrate the effectiveness of the proposed Re4 optimization method. Empirically, CBR+Re4 outperforms other baselines on datasets collected from two product development units (PDUs) at Huawei Datacom. Notably, CBR+Re4 can also effectively alleviate the repetitive generation issues encountered with the previously deployed CBR+SFT method.

We summarize the contributions of our work as follows:
\begin{itemize}
    \item To the best of our knowledge, this is the first attempt to utilize LLMs to assist test engineers in functional testing.
    \item We propose a CBR system to enhance the capabilities of LLMs for functional test script generation. 
    \item We introduce the Re4 method to finetune both the retriever model and the LLM within the CBR system, ensuring better alignment with our production scenario.
    \item We conduct extensive experiments on real-world datasets from Huawei Datacom to demonstrate the superiority of the proposed Re4 method. Meanwhile, our findings show that CBR+Re4 alleviates the repetitive generation issues of LLMs, further improving the user experience.
\end{itemize}

\section{Preliminaries}
To clearly explain the problem under investigation, we first introduce the overall case-based reasoning framework for functional test script generation with large language models, and then present the business metric in our production scenario.

\subsection{The Case-Based Reasoning System}
In this work, we focus on investigation of functional test script generation with large language models (LLM). Given a test intent description $q$, LLMs are tasked with generating the corresponding test script $y$, which can be framed as a text-to-code generation problem. However, different from general text-to-code generation problems and other test script generation tasks (e.g., unit test generation), functional test script generation requires LLMs to understand detailed code structures of the target software. Given that the size of industrial software typically exceeds the token capacity of LLMs, this task still remains underexplored \cite{llm4test-survey}. 

In our production scenario, there exist a large collection of functional test scripts written by human test engineers for each software, containing rich implicit knowledge that maps test intent description to test scripts. To fully harness this wealth of knowledge, we adopt a classic problem-solving paradigm, case-based reasoning (CBR) \cite{cbr-review-1, cbr-review-2, 4R-cycle}, to enhance the capabilities of LLMs for functional test script generation. Specifically, we follow the classic 4R CBR cycle \cite{4R-cycle}, which contains four steps: \textbf{(1) Retrieve} similar cases from the case bank; \textbf{(2) Reuse} the retrieved cases to solve the current task; \textbf{(3) Revise} the solution to guarantee the correctness for the current task; \textbf{(4) Retain} the task and the solution into the case bank. The overall workflow of CBR system is demonstrated in Figure \ref{fig:framework}(a).

We now provide a detailed description of the CBR system for functional test script generation with LLMs. Given a test intent description $q$, the CBR system first retrieves top-$M$ similar cases from the case bank $\mathcal{C}=\{c_i\}_{i=1}^{N}$ with each case $c_i=(q_i,y_i)$. To achieve this, we utilize a pretrained embedding model $\mathbf{E}_\phi(\cdot)$ as the retriever model to calculate the semantic similarity between two test intent descriptions $q$ and $q'$ as: 
\begin{equation}
    \label{eq:cos-sim}   \text{sim}_\phi(q,q')=\cos\langle\mathbf{E}_\phi(q),\mathbf{E}_\phi(q')\rangle.
\end{equation}
Next, the LLM $\pi_\theta(\cdot)$ reuses the retrieved cases $\mathcal{C}_q=\{c_i\}_{i=1}^{M}$ to generate the test script, i.e., $\hat{y}\sim \pi_\theta(\cdot|q,\mathcal{C}_q)$. Human test engineers then verify the correctness of the generated test scripts and revise them as necessary. Finally, the test intent description $q$ and the corresponding revised test script $y$ are retained into the case bank, i.e., $\mathcal{C}\leftarrow\mathcal{C}\cup\{(q,y)\}$. We summarize the workflow of this CBR system in Algorithm \ref{alg:cbr-inference} in Appendix \ref{app:algorithm}.

Note that the aforementioned Retrieve-Reuse process in the CBR system is similar to the retrieval-augmented generation (RAG) techniques \cite{rag-1, rag-2, rag-survey}. The key difference is that RAG retrieves relevant document chunks to provide detailed context for LLM generation, whereas CBR retrieves similar cases to enable analogical reasoning for task-solving. Furthermore, the CBR system benefits from the Retain step to achieve a flexible learning mechanism during deployment, eliminating the need for resource-intensive finetuning. 

\subsection{Business Metric}
\label{sec:metric}
In the context of our production scenario, the goal is to assist human test engineers in improving their efficiency in writing functional test scripts, rather than fully automating the process in current stage. This decision is driven by two main considerations: (1) the provided test intent descriptions are often not detailed enough to enable complete test script generation by LLMs; (2) the retrieved cases may not comprehensively cover the software structures required for functional test script generation. Therefore, the acceptance rate, the percentage of generated test scripts accepted by human test engineers, can be regarded as the online business metric. However, this metric is significantly biased in practice, as user preferences for accepting generated test scripts vary widely. Worse still, the limited number of users within each PDU prevents the mitigation of bias through the law of large numbers.

To solve this issue, we design offline business metrics by evaluating the script similarity between the generated test script $\hat{y}$ and the ground-truth test script $y$. This task is particularly challenging due to the vast space of functionally equivalent code. One notable characteristic of our production scenario is that functional test scripts focus on invoking functions to structure the desired workflow. Thus, the accuracy of functions within the generated test script is crucial for achieving high user satisfaction.
Based on this observation, we propose to measure the script similarity with three offline business metrics: function precision, function recall and function f1 score, defined as follows:
\begin{align}
    \text{FPrecision}(\hat{y},y)&=\frac{|\text{Func}(\hat{y})\cap\text{Func}(y)|}{|\text{Func}(\hat{y})|}, \label{eq:precision} \\
    \text{FRecall}(\hat{y},y)&=\frac{|\text{Func}(\hat{y})\cap\text{Func}(y)|}{|\text{Func}(y)|}, \label{eq:recall}\\
    \text{FF1}(\hat{y},y)&=\frac{2\cdot\text{FPrecision}(\hat{y},y)\cdot\text{FRecall}(\hat{y},y)}{\text{FPrecision}(\hat{y},y)+\text{FRecall}(\hat{y},y)}, \label{eq:f1}
\end{align}
where $\text{Func}(\cdot)$ returns the set of functions invoked in the test script. Among them, function precision measures the accuracy of the functions called in the generated test script, while function recall evaluates how well the generated test script covers the functions in the ground-truth test script. The function f1 score balances both precision and recall, providing a comprehensive measure of overall performance. Hence, the function f1 score is the most crucial business metric in our production scenario.

In addition to these three metrics, we also incorporate code similarity \cite{code-similarity}, measured using a normalized Levenshtein distance \cite{levenshtein-distance} score, as a complementary metric. We defer implementation details of the aforementioned offline business metrics to Appendix~\ref{app:metric}.
\section{Methodology}
In this section, we aim to optimize the aforementioned CBR system for better alignment with our production scenario. As Revise and Retain steps do not involve machine learning models, we focus on optimizing the retriever model $\mathbf{E}_\phi(\cdot)$ in the Retrieve step, and the LLM $\pi_\theta(\cdot)$ in the Reuse step. We introduce Re4, an optimization method for our CBR system, comprising two key components: (1) a reranking-based retrieval finetuning method, which optimizes the retriever model to retrieve cases more similar to the query, and (2) a reinforced reuse finetuning method, which enables LLMs to faithfully reuse the retrieved cases for solving new tasks. The overall framework of Re4 is illustrated in Figure \ref{fig:framework}.

In this work, we adopt the offline training strategy to avoid high computational costs required in real-time online training. Let $\mathcal{D}=\{(q_i,y_i)\}_{i=1}^{N}$ be the training set with $N$ samples, where $q_i$ is the $i$-th test intent description, and $y_i$ is the corresponding test script. For each sample $(q_i,y_i)\in\mathcal{D}$, we regard the other samples in the training set as the case bank, i.e., $\mathcal{C}=\mathcal{D}\setminus \{(q_i,y_i)\}$.

\subsection{Reranking-based Retrieval Finetuning}
In the CBR system, the Retrieve step is crucial for retrieving similar cases from the case bank to support the subsequent Reuse step. While pretrained embedding models offer strong zero-shot retrieval performance, we aim to further finetune the models to better align with the specific corpus of our production scenario. Therefore, we utilize contrastive learning, a widely adopted approach for representation learning, to finetune the pretrained embedding model. Formally, given a test intent description $q$ with an associated positive example $c^+=(q^+,y^+)$ and a pool of negative examples $\mathcal{C}^-$, the well-known InfoNCE loss \cite{infoNCE, retrieval-infoNCE} is defined as 
\begin{equation}
    \label{eq:infonce}
    \mathcal{L}(\phi)=-\frac{\exp(\text{sim}_\phi(q,q^+)/\tau)}{\exp(\text{sim}_\phi(q,q^+)/\tau)+\sum_{q^-\in\mathcal{C}^-}{\exp(\text{sim}_\phi(q,q^-)/\tau)}},
\end{equation}
where $\tau$ denotes the hyper-parameter for temperature. Different from general retrieval problems \cite{supervised-retrieval-1, supervised-retrieval-2, supervised-retrieval-3, supervised-retrieval-4}, positive examples are not explicitly available in our setting, making the key challenge the effective mining of positive and negative examples for a given query. One promising approach is to leverage LLMs to provide pseudo-labels of positive and negative examples \cite{rankgpt, replug, dpa-rag}. However, this requires multiple rounds of sampling from LLMs, resulting in significant computational and time costs.

To address this challenge, we aim to identify the positive example based on the nature of our production scenario. As the basic assumption of CBR \cite{cbr-assumption-1, cbr-assumption-2} — that similar problems have similar solutions — is invertible in text-to-code generation due to its translation nature, we can expect that similar test scripts should correspond to similar test intent descriptions.
Therefore, we identify positive examples as those with both high semantic similarity and script similarity.
Since functional testing emphasizes invoking functions to construct workflows, the script similarity can be effectively measured using the function f1 score, as defined in Eq. (\ref{eq:f1}). Based on the aforementioned findings, we propose a reranking-based retrieval finetuning method, as shown in Figure \ref{fig:framework}(b).

Given a case $(q,y)$, we first retrieve top-$k$ cases from the case bank $\mathcal{C}$ based on the semantic similarity defined in Eq. (\ref{eq:cos-sim}), i.e., 
\begin{equation}
\label{eq:retrieve}
\mathcal{C}'=\underset{(q',y')\in\mathcal{C}}{\arg\text{top-}k}~\text{sim}_\phi(q,q').
\end{equation}
Then, we rerank the retrieved cases $\mathcal{C}'$ based on the script similarity with function f1 score. The positive case $c^+$ is labeled as the case with most similar test script, and the remaining cases form the pool of negative cases $\mathcal{C}^-$, as follows:
\begin{align}
\label{eq:pos}
c^+&=\underset{(q',y')\in\mathcal{C}'}{\arg\max}\text{FF1}(y',y), \\
\label{eq:neg}
\mathcal{C}^-&=\mathcal{C}'\setminus\{c^+\}.
\end{align}
As such, we can follow the InfoNCE loss defined in Eq. (\ref{eq:infonce}) to finetune the pretrained embedding model. Following previous works \cite{supervised-retrieval-1, supervised-retrieval-2, in-batch-negative}, we also include in-batch negative examples for the InfoNCE loss, which has been shown to be an effective trick that boosts the number of training examples. We summarize the reranking-based retrieval finetuning method in Algorithm \ref{alg:retrieval} in Appendix \ref{app:algorithm}.

\subsection{Reinforced Reuse Finetuning}
In the CBR system, the Reuse step focuses on adapting solutions of past similar cases to the current task. While modern LLMs demonstrate strong instruction-following capabilities, they may still suffer from misalignment issues \cite{POAD,TWOSOME,RL-LLM-Prior} when confronted with unseen tasks. Therefore, further finetuning of LLMs is necessary to ensure alignment with the desired behavior in our production scenarios. To this end, we propose a reinforced reuse finetuning method that incorporates both supervised finetuning (SFT) and reinforcement learning finetuning (RLFT), as shown in Figure \ref{fig:framework}(c).

\subsubsection{Supervised Finetuning}
\label{sec:sft}
Given a training sample $(q,y)\in\mathcal{D}$ and its corresponding retrieved cases $\mathcal{C}_q$, we can perform standard SFT by maximizing the log probability of each token in $y$. The loss function of SFT can be formulated as
\begin{equation}
    \label{eq:sft}
    \mathcal{L}_{\text{SFT}}(\theta)=-\log\pi_\theta(y|q,\mathcal{C}_q).
\end{equation}
Then, we can derive a finetuned LLM with parameters $\theta_{\text{SFT}}$, which is also the starting point for RLFT.

\begin{figure}
    \includegraphics[width=0.4\textwidth]{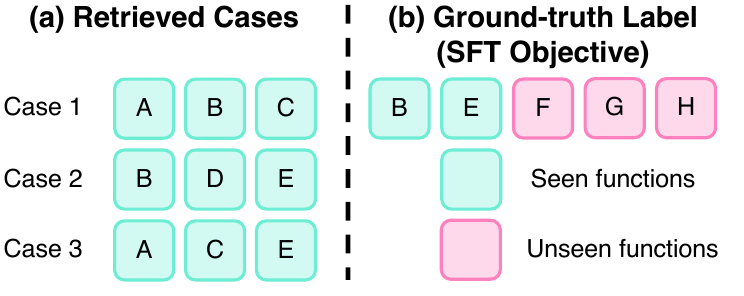}
    \caption{A motivating example for the limitation of SFT. SFT objective may contain unseen functions beyond the retrieved cases, resulting in noisy alignment.}
    \label{fig:motivation}
    \vspace{-10pt}
\end{figure}

\subsubsection{Reinforcement Learning Finetuning} While SFT is efficient for alignment, it can be problematic in our production scenario, which requires LLMs precisely invoking functions contained in the retrieved cases to structure the testing workflow. Now, we present a motivating example. As illustrated in Figure \ref{fig:motivation}, the retrieved cases include five functions (A, B, C, D, E), whereas the ground-truth label involves five different functions (B, E, F, G, H), three of which (F, G, H) are absent from the retrieved cases. Since the SFT objective enforces fitting the ground-truth label, it compels LLMs to generate unseen functions, introducing noise into alignment and exacerbating hallucination issues during inference. To address this issue, we introduce an online RLFT stage to further refine the alignment and mitigate these negative effects. Specifically, given a sample $(q,y)\in\mathcal{D}$ and its corresponding retrieved cases $\mathcal{C}_q=\{(q_i,y_i)\}_{i=1}^M$, the optimization objective of RLFT can be formulated as:
\begin{equation}
    \max_\theta \mathbb{E}_{\hat{y}\sim \pi_\theta(\cdot|q,\mathcal{C}_q)}[r(\hat{y})],
\end{equation}
where $r(\cdot)$ is the reward function that quantifies the quality of the generated script. Here, we utilize the script similarity between the generated script $\hat{y}$ and ground-truth label $y$ as the golden reward, which can be effectively measured by the function f1 score defined in Eq. (\ref{eq:f1}). Thus, we can reformulate the optimization objective as:
\begin{equation}
    \label{eq:final-oo}
    \max_\theta \mathbb{E}_{\hat{y}\sim \pi_\theta(\cdot|q,\mathcal{C}_q)}[\text{FF1}(\hat{y},y)-\beta \cdot \mathbb{D}_{\text{KL}}(\pi_\theta||\pi_{\theta_\text{SFT}})],
\end{equation}
where we follow previous works \cite{rloo,remax,grpo,TDPO} to incorporate a KL divergence penalty term to avoid too much deviation from the reference SFT policy model $\pi_{\theta_\text{SFT}}$, and $\beta$ denotes the coefficient for the KL divergence penalty. As such, we can encourage faithful reuse that invokes the correct functions, while penalizing the inclusion of incorrect functions resulting from the hallucination issue. Meanwhile, this objective does not push LLMs to fit unseen functions as SFT, thereby avoiding the introduction of new noise during RLFT.

To solve the problem defined in Eq. (\ref{eq:final-oo}), proximal policy optimization (PPO) \cite{ppo} is the most widely adopted algorithm; however, it is overly complicated for optimization and hyper-parameter tuning, especially for LLMs. To avoid unnecessary complicated designs, many previous works, such as RLOO \cite{rloo}, Remax \cite{remax} and GRPO \cite{grpo}, devote to the simplification of PPO algorithm. However, they typically require two or more times of on-policy sampling for variance reduction in the optimization process, leading to additional computational and time costs and limiting the application in our production scenario. Different from these algorithms designed for reinforcement learning from human feedback (RLHF) for open-ended text generation, our RLFT setting requires LLMs to focus on the reuse of the retrieved cases for test script generation, significantly narrowing the action space. As such, the randomness from the sampling of LLMs is far less compared to open-ended text generation, thereby weakening the high variance issue of the stochastic gradient. This enables us to omit all the variance reduction techniques and back to the most basic on-policy RL algorithm REINFORCE \cite{reinforce} for RLFT, which can be formulated as:
\begin{equation}
    \label{eq:reinforce}
    \mathcal{L}_{\text{REINFORCE}}(\theta)=\mathbb{E}_{\hat{y}\sim \pi_\theta(\cdot|q,\mathcal{C}_q)}[-r(\hat{y})\cdot\log\pi_\theta(\hat{y}|q,\mathcal{C}_q))],
\end{equation}
where $r(\hat{y})$ denotes the reward function containing the function f1 score and the KL divergence penalty term as in Eq.~(\ref{eq:final-oo}). We provide a more detailed discussion on why REINFORCE works in our RLFT setting in Appendix~\ref{app:reinforce-works}.
We summarize the reinforced reuse finetuning method in Algorithm \ref{alg:reuse} in Appendix~\ref{app:algorithm}.

\begin{table*}[htbp]
\caption{Performance comparison of different methods across two datasets. The best results are highlighted in bold. We also report the relative improvement of CBR+Re4 over the best baseline.}
\label{exp:main}
\begin{tabular}{@{}lccccccccc@{}}
\toprule
 & \multicolumn{4}{c}{\textbf{DCN}} & \textbf{} & \multicolumn{4}{c}{\textbf{SP}} \\ \cmidrule(lr){2-5} \cmidrule(l){7-10} 
 & \textbf{CS ($\uparrow$)} & \textbf{FP ($\uparrow$)} & \textbf{FR ($\uparrow$)} & \textbf{FF1 ($\uparrow$)} & \textbf{} & \textbf{CS ($\uparrow$)} & \textbf{FP ($\uparrow$)} & \textbf{FR ($\uparrow$)} & \textbf{FF1 ($\uparrow$)} \\ \midrule
\textbf{Naïve SFT} & 47.62 & 57.28 & 50.40 & 49.46 &  & 43.72 & 42.35 & 39.50 & 38.71 \\
\textbf{Naïve CBR} & 53.63 & 67.11 & 67.93 & 64.01 &  & 56.47 & 55.97 & 57.09 & 54.60 \\
\textbf{CBR} & 54.80 & 66.47 & 70.13 & 64.67 &  & 54.49 & 56.52 & 59.78 & 55.83 \\
\textbf{CBR+SFT} & 63.38 & 74.09 & 68.89 & 67.87 &  & 61.10 & 61.31 & 58.27 & 57.98 \\
\textbf{CBR+Re4 (Ours)} & \textbf{64.60} & \textbf{75.50} & \textbf{71.33} & \textbf{70.11} & \textbf{} & \textbf{62.38} & \textbf{64.89} & \textbf{62.43} & \textbf{61.92} \\ \midrule
\textbf{Improvement} & \textbf{+1.92\%} & \textbf{+1.90\%} & \textbf{+3.54\%} & \textbf{+3.30\%} & \textbf{} & \textbf{+2.09\%} & \textbf{+5.84\%} & \textbf{+7.14\%} & \textbf{+6.80\%} \\ \bottomrule
\end{tabular}
\end{table*}

\section{Experiments}
\subsection{Experimental Setups}
\subsubsection{Dataset} We collect datasets from two product development units (PDU) in Huawei Datacom, referred to as Data Communication Network (DCN) and Software Platform (SP). The DCN dataset consists of 30,887 samples, while the SP dataset contains 58,429 samples. The collected datasets include noisy samples, such as those with poor coding practices and ambiguous or lengthy descriptions. To minimize human effort, we opt not to pre-process them further. Instead, we curate a clean and representative testing dataset to ensure reliable evaluation, comprising 366 samples for SP and 689 samples for DCN. We use consistent train-validation-test splits across all the methods.

\subsubsection{Experiment Setting} The CBR system in this paper involves two backbone models, a pretrained embedding model for the Retrieve step and a LLM for the Reuse step. We utilize bge-m3 \cite{bge-m3} as the embedding model, and an internal LLM with approximately 7B parameters as the base model across all the methods. Due to computational requirements, we utilize LoRA \cite{lora} for finetuning, enabling training to be conducted on a single Huawei Ascend 910B NPU with 64GB of memory.

For the CBR system, we retrieve $M=3$ cases from the case bank for both training and inference stage. For the optimization of the Retrieve step, we finetune the embedding model for five epochs, setting the temperature $\tau$ to 1.0, learning rate to 1e-6, and batch size to 64 for both datasets. In terms of the optimization for the Reuse step, we perform SFT for one epoch with the batch size of 32. As for RLFT, we finetune the LLM with one epoch and batch size of 64, using a KL divergence coefficient $\beta$ of 0.1. For both SFT and RLFT, we set the learning rate as 3e-5 for DCN and 1e-5 for SP. Additionally, we apply a cosine scheduler with a 3\% warmup for the learning rate.

During training, we set the LLM's sampling temperature to 0.9 to enhance the diversity of the generated test scripts. During inference, we adopt greedy decoding to exclude randomness.

\subsubsection{Evaluation Metric} During evaluation, we regard the training set as the case bank. We adopt four evaluation metrics: code similarity (CS), function precision (FP), function recall (FR), and function F1 score (FF1), as detailed in Section \ref{sec:metric}. Among these, FF1 score is the most critical metric for evaluation.
 
\subsubsection{Baselines} To the best of our knowledge, this is the first paper that discusses the optimization of the full workflow in the CBR system with LLMs. Thus, there are not any other previous works suitable for comparison. To contextualize the performance of the proposed method, we carefully design four baselines:
\begin{itemize} 
    \item \textbf{Naïve SFT}: It directly performs supervised finetuning for test script generation without applying CBR techniques.
    \item \textbf{Naïve CBR}: It implements the simplest form of CBR by retrieving a single case from the case bank as the generated test script. Despite its simplicity, this approach is a strong baseline in our early attempts.
    \item \textbf{CBR}: It retrieves three similar cases from the case bank and uses them as context to prompt the LLM for generating test scripts. It does not perform any finetuning for both retriever model and LLM.
    \item \textbf{CBR+SFT}: Built on top of CBR, it further applies SFT for LLMs with three retrieved cases in the context, as described in Section \ref{sec:sft}. Note that it does not perform finetuning for the retriever model.
\end{itemize}

\begin{figure}
    \includegraphics[width=0.5\textwidth]{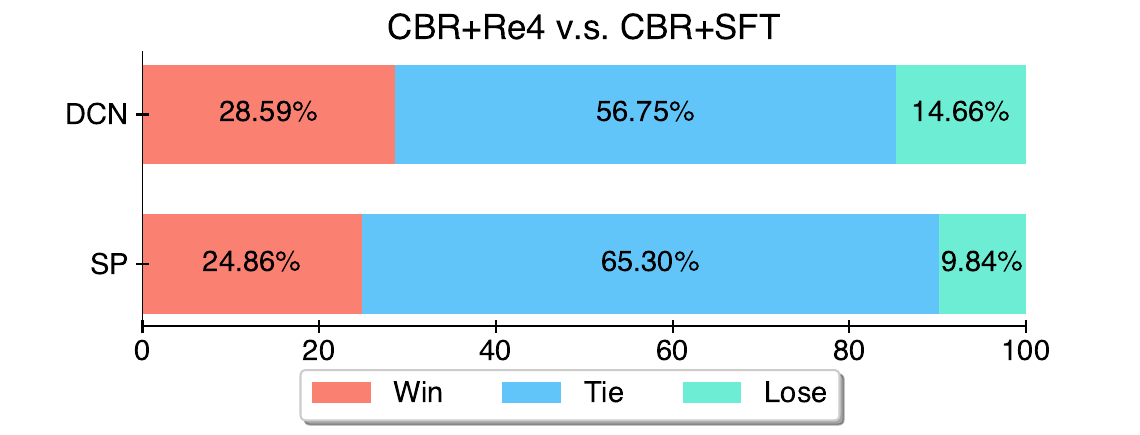}
    \caption{Comparison between CBR+Re4 and CBR+SFT. The win, tie and lose rates are evaluated by humans.}
    \label{fig:win-tie-lose}
    \vspace{-10pt}
\end{figure}

\subsection{Main Results}
\subsubsection{Overall Comparison} We present the experimental results of the offline evaluation metrics for both datasets in Table \ref{exp:main}. Note that we follow most LLM works to report the result of a single run due to the overly computational costs, which might be one limitation of this work. We can observe that Naïve SFT performs the worst among all methods, even underperforming the simplest baseline, Naïve CBR. This is expected, as functional test script generation requires the LLM to invoke functions beyond its knowledge, which cannot be directly injected by SFT. Consequently, this leads to significant hallucination issues during inference. In contrast, CBR outperforms Naïve CBR, thanks to the strong foundational capabilities of LLMs. These capabilities are further enhanced by incorporating SFT, making CBR+SFT the strongest baseline. Among all the methods, the proposed CBR+Re4 achieves the highest FF1 score for both datasets, showing improvement of 3.30\% and 6.80\% over the best baseline, CBR+SFT, for DCN and SP, respectively. This validates the effectiveness of both the reranking based retrieval finetuning method and the reinforced reuse finetuning method.

\subsubsection{Human Evaluation} 
\label{sec:human-evaluation}
To comprehensively evaluate the quality of the generated test scripts, we perform pairwise human evaluations comparing the proposed CBR+Re4 with the best baseline, CBR+SFT, on two datasets. The win, tie, and loss rates are reported in Figure \ref{fig:win-tie-lose}. The results show that CBR+Re4 achieves a higher win rate on both datasets compared to CBR+SFT, further validating the superiority of the proposed method.

\begin{table*}[htbp]
\caption{In-depth analyses of the proposed reinforced reuse finetuning method across two datasets. We highlight those results better than CBR+Re4 with \textcolor{white}{\colorbox[HTML]{FFEBD6}{\hspace{0.5em}\rule{0pt}{0.5em}\hspace{0.5em}}}.}
\label{exp:reuse}
\begin{tabular}{@{}lcccccccccc@{}}
\toprule
 &  & \multicolumn{4}{c}{\textbf{DCN}} & \textbf{} & \multicolumn{4}{c}{\textbf{SP}} \\ \cmidrule(lr){3-6} \cmidrule(l){8-11} 
 & \multirow{-2}{*}{\textbf{\#On-policy samples}} & \multicolumn{1}{l}{\textbf{CS ($\uparrow$)}} & \multicolumn{1}{l}{\textbf{FP ($\uparrow$)}} & \multicolumn{1}{l}{\textbf{FR ($\uparrow$)}} & \multicolumn{1}{l}{\textbf{FF1 ($\uparrow$)}} & \textbf{} & \multicolumn{1}{l}{\textbf{CS ($\uparrow$)}} & \multicolumn{1}{l}{\textbf{FP ($\uparrow$)}} & \multicolumn{1}{l}{\textbf{FR ($\uparrow$)}} & \multicolumn{1}{l}{\textbf{FF1 ($\uparrow$)}} \\ \midrule
\rowcolor[HTML]{EFEFEF} 
\textbf{CBR+Re4 (Ours)} & 1 & 64.60 & 75.50 & 71.33 & 70.11 &  & 62.38 & 64.89 & 62.43 & 61.92 \\ \midrule
\multicolumn{11}{c}{\textit{Ablation study for reinforced reuse finetuning method}} \\ \midrule
\textbf{CBR+Re4 w/o Finetuning} & 0 & 55.39 & 68.81 & 71.03 & 66.06 &  & 57.72 & 59.83 & \cellcolor[HTML]{FFEBD6}63.01 & 59.17 \\
\textbf{CBR+Re4 w/o SFT} & 1 & 55.24 & 69.87 & 70.94 & 66.60 &  & 58.78 & 61.02 & \cellcolor[HTML]{FFEBD6}63.12 & 60.02 \\
\textbf{CBR+Re4 w/o RLFT} & 0 & 64.10 & 74.8 & 69.44 & 68.64 &  & 60.92 & 61.99 & 59.29 & 59.06 \\ \midrule
\multicolumn{11}{c}{\textit{Comparison with different RL algorithms in reinforced reuse finetuning method}} \\ \midrule
\textbf{CBR+Re4 w/ Online DPO} & 2 & 63.99 & 74.76 & 69.31 & 68.58 & \textbf{} & 60.71 & 61.98 & 59.56 & 59.09 \\
\textbf{CBR+Re4 w/ Remax} & 2 & 64.35 & 75.03 & 69.46 & 68.80 &  & 60.99 & 62.02 & 59.36 & 59.10 \\
\textbf{CBR+Re4 w/ RLOO} & 4 & 64.76 & 75.62 & 71.01 & 69.88 &  & 61.87 & 63.44 & 62.05 & 60.97 \\
\textbf{CBR+Re4 w/ GRPO} & 4 & \cellcolor[HTML]{FFEBD6}65.67 & \cellcolor[HTML]{FFEBD6}75.69 & \cellcolor[HTML]{FFEBD6}72.29 & \cellcolor[HTML]{FFEBD6}70.80 &  & \cellcolor[HTML]{FFEBD6}63.18 & 64.82 & \cellcolor[HTML]{FFEBD6}62.83 & \cellcolor[HTML]{FFEBD6}62.17 \\ \bottomrule
\end{tabular}
\end{table*}

\begin{figure}[tbp]
    \includegraphics[width=0.5\textwidth]{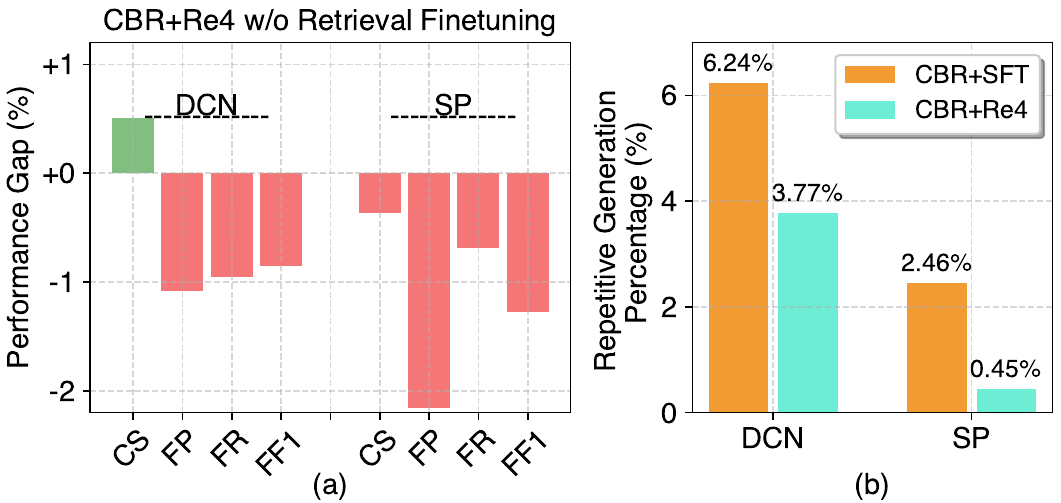}
    \caption{(a) Performance gap in an ablation study of CBR+Re4 w/o retrieval finetuning. (b) Repetitive generation percentage of different methods.}
    \label{fig:retrieval-ablation-repetitive-generation}
    \vspace{-10pt}
\end{figure}

\subsection{In-Depth Analyses for Re4}
In this subsection, we present in-depth analyses for the proposed finetuning methods for both retriever model and LLM.
\subsubsection{Analyses for Reranking-based Retrieval Finetuning} As the key challenge for finetuning retriever models is lack of ground-truth labels, most advanced finetuning methods cannot be adapted to our setting. Thus, we analyze the proposed reranking-based retrieval finetuning method via an ablation study. The performance gap of CBR+Re4 without retrieval finetuning is reported for both datasets in Figure~\ref{fig:retrieval-ablation-repetitive-generation}(a). The results show that the ablation causes a performance drop of approximately 1\% in terms of FF1 score on both datasets. This highlights the effectiveness of the proposed reranking-based retrieval finetuning method.

\subsubsection{Analyses for Reinforced Reuse Finetuning} Now, we present in-depth analyses for the proposed reinforced reuse finetuning method. First, we conduct an ablation study to validate its effectiveness. Next, we replace the proposed REINFORCE algorithm with several state-of-the-art RLHF algorithms to highlight the superiority of our approach. We then demonstrate an additional advantage of the RLFT stage in mitigating repetitive generation issues. Finally, we analyze the impact of the KL coefficient $\beta$ through a hyper-parameter analysis.

\noindent\textbf{Ablation study.} We first conduct an ablation study of CBR+Re4 by evaluating the following ablation variants: \textbf{(1) w/o Finetuning}, which relies solely on the foundational capabilities of the LLM; \textbf{(2) w/o SFT}, which applies RLFT directly to the base LLM without the prior SFT stage; and \textbf{(3) w/o RLFT}, which involves only SFT stage without the subsequent RLFT stage.

The corresponding results are presented in the upper section of Table~\ref{exp:reuse}. Among them, CBR+Re4 w/o Finetuning demonstrates the poorest performance on both datasets, emphasizing the necessity of further aligning the LLM with our production scenario. Additionally, both CBR+Re4 w/o SFT and w/o RLFT underperform CBR+Re4. This outcome aligns with expectations: SFT may introduce undesired behaviors due to the noise in ground-truth labels, while RLFT may suffer from sample inefficiency. The combined SFT-RLFT paradigm, consistent with best practices in the RLHF community, strikes the desired trade-off between effectivenss and efficiency.

\noindent\textbf{Comparison with advanced RLHF algorithms.} We compare the proposed REINFORCE algorithm with several state-of-the-art RLHF algorithms. Unlike typical RLHF settings, our production scenario leverages a golden reward function instead of a trained reward model. Moreover, we focus on improving the case-based reasoning capabilities instead of general instruction following as in RLHF. Specifically, we evaluate the following algorithms:
\textbf{(1) Online DPO} \cite{online-dpo}, which performs pairwise comparisons with the reward function and optimizes the LLM using DPO \cite{dpo} in an online manner.  
\textbf{(2) Remax} \cite{remax}, which incorporates a baseline via greedy decoding in the REINFORCE algorithm to reduce variance.  
\textbf{(3) RLOO} \cite{rloo}, which employs a variance-reduced multi-sample estimate for policy updates.  
\textbf{(4) GRPO} \cite{grpo}, which calculates the advantage in the PPO objective based on group-level relative rewards.  
Due to space limitations, implementation details are provided in Appendix~\ref{app:rlhf}. Note that we exclude the PPO \cite{ppo} algorithm from our comparison due to its high computational costs. Furthermore, prior RLHF works have demonstrated that Remax, RLOO, and GRPO achieve superior alignment compared to PPO. 

The experimental results are presented in the lower section of Table~\ref{exp:reuse}. These results show that all four RLHF algorithms further improve alignment compared to the SFT stage, confirming the necessity of the RLFT stage. Additionally, the proposed REINFORCE algorithm outperforms Online DPO, Remax, and RLOO, while delivering competitive performance relative to GRPO, further demonstrating its effectiveness. Notably, all these RLHF algorithms require two or more on-policy samples per query, significantly increasing time and computational costs. In contrast, the proposed REINFORCE algorithm requires only a single on-policy sample per query, striking the desired trade-off between the alignment performance and costs of on-policy sampling.

\begin{figure}[tbp]
    \includegraphics[width=0.45\textwidth]{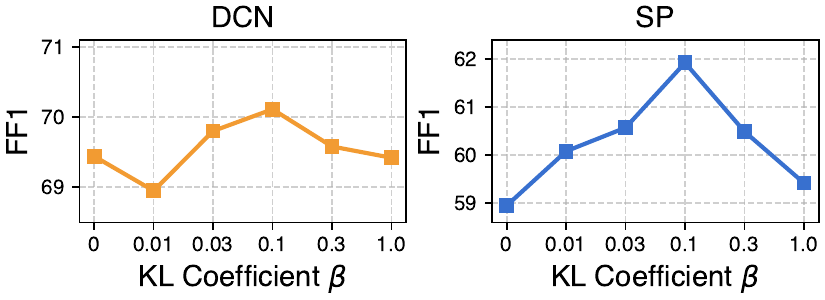}
    \caption{Hyper-parameter analyses of KL coefficient $\beta$ on both datasets.}
    \label{fig:kl-coeeficient}
    \vspace{-10pt}
\end{figure}

\noindent\textbf{Analysis on repetitive generation issue.} During prior online deployment of CBR+SFT, we observed that the generated test scripts often exhibited the repetitive generation issue \cite{repetitive-1, repetitive-2}. An example of such repetitive generation is provided in Figure~\ref{fig:repetitive-example} in Appendix~\ref{app:rgi}. This issue significantly slows response times, increases inference costs, and adversely impacts user experience. As such patterns are challenging to detect using rule-based methods, we conduct a human evaluation to assess the repetitive generation issue in the test scripts generated by CBR+SFT and CBR+Re4.

The percentage of repetitive generation is presented in Figure~\ref{fig:retrieval-ablation-repetitive-generation}(b). As observed, CBR+Re4 effectively mitigates the repetitive generation issue, reducing it by 2.47\% for DCN and 2.41\% for SP. Notably, CBR+Re4 exhibits only 0.45\% repetitive generation in SP, highlighting the superiority of the RLFT stage. This improvement is understandable, as SFT may memorize poor coding practices from the training corpus, while the noise in the ground-truth labels further exacerbates these undesired behaviors. In contrast, during the RLFT stage, test scripts with repetitive generation patterns receive near-zero rewards due to repetitive invocation of the same functions. As such, these repetitive behaviors are penalized by the on-policy RL objective, thereby alleviating the issue.

\noindent\textbf{Hyper-parameter analysis on KL coefficient $\beta$.} Finally, we analyze an important hyperparameter in the reinforced reuse finetuning method: the KL divergence coefficient $\beta$. We evaluate $\beta$ at values of \{0, 0.01, 0.03, 0.1, 0.3, 1.0\} and report the FF1 score on both datasets in Figure~\ref{fig:kl-coeeficient}. Notably, $\beta=0$ serves as an ablation study of the KL divergence term in Eq.~(\ref{eq:reinforce}), resulting in a significant performance decline. This highlights the importance of preventing too much deviation from the finetuned LLM parameters. As expected, setting $\beta$ too conservatively or too aggressively can negatively impact performance. As shown in Figure~\ref{fig:kl-coeeficient}, $\beta=0.1$ yields the best performance on both datasets.

\begin{figure}[tbp]
    \includegraphics[width=0.5\textwidth]{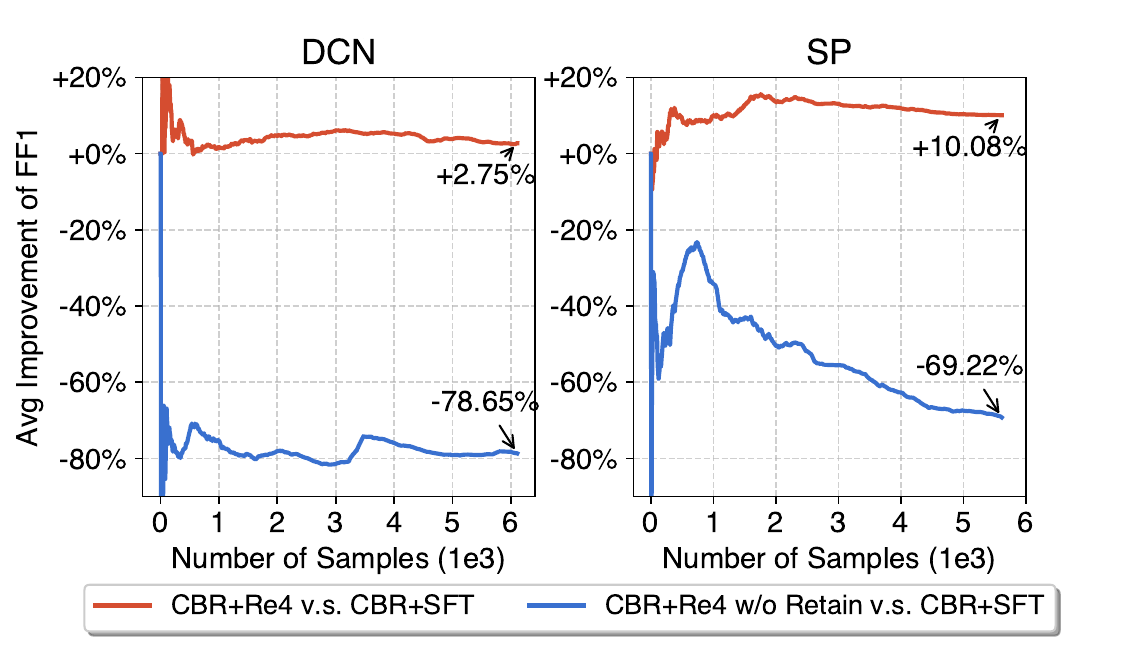}
    \caption{Online evaluation results of CBR+Re4 on two PDUs.}
    \label{fig:online}
    \vspace{-10pt}
\end{figure}

\subsection{Online Evaluation Results}
In this subsection, we provide an online evaluation of the proposed CBR system. A natural approach to quantifying online performance is to estimate the acceptance rate through online A/B tests. However, this can be problematic in our production scenario. Specifically, users exhibit significant variability in their preferences for accepting the generated test scripts, resulting in biased evaluation outcomes. Worse still, the limited number of users within each PDU prevents the law of large numbers from eliminating this bias. In addition, the human evaluation method described in Section~\ref{sec:human-evaluation} is prohibitively expensive for assessing online performance. To address this challenge, we report the online performance with FF1 score, evaluated on approximately 6,000 recent online samples from the deployed system across two PDUs. The evaluation aligns with the workflow of the CBR system: sequentially processing the user requests of test intent description, generating test scripts with the Retrieve and Reuse step, and finally retaining the test intent description with the revised test script into the case bank.

We compare the previously deployed \textbf{CBR+SFT} system with two methods: \textbf{(1) CBR+Re4} and \textbf{(2) CBR+Re4 w/o Retain}, an ablation variant of the CBR system where the Retain step is removed and the case bank remains fixed. It is worth noting that CBR+Re4 w/o Retain can be viewed as a variant of RAG, as it only reserves the Retrieve-Reuse process in CBR. Figure~\ref{fig:online} depicts the average improvement in FF1 score as the number of online samples increases across two PDUs, DCN and SP. We observe that CBR+Re4 w/o Retain significantly underperforms CBR+Re4. This performance gap can be attributed to the dynamic nature of our scenario, where the software is continuously updated, and new modules are regularly introduced. Consequently, updating the case bank in an online manner is crucial for maintaining the effectiveness of test script generation. The Retain step in the CBR system addresses this need by enabling seamless integration of new information into the case bank. Furthermore, the proposed CBR+Re4 consistently outperforms the previously deployed CBR+SFT, achieving improvements in FF1 score of 2.75\% in DCN and 10.08\% in SP. These results validate the effectiveness of the proposed Re4 optimization method. 
\section{Related Work}
\subsection{LLMs for Software Testing}
LLMs have demonstrated remarkable success in the field of code generation \cite{DS-Agent,agentk,debugbench,swe-bench,mftcoder,unitrans}. Recently, the software testing community has been exploring the potential of LLMs for automated software testing \cite{llm4test-survey}, with a primary focus on unit test generation. For instance, Wang et al. \cite{test-eval} introduce TestEval, a benchmark designed to evaluate the capabilities of LLMs in generating unit test cases. Additionally, Alshahwan et al. \cite{meta-unit-test} develop an autonomous workflow at Meta, enabling LLMs to improve assured unit test cases without human intervention. Different from these prior works, we explore the application of LLMs for functional test script generation, which necessitates a deep understanding of the complex code structure of the target software.

\subsection{Case-Based Reasoning}
CBR \cite{cbr-review-1,cbr-review-2,4R-cycle} is a classic AI paradigm that operates by retrieving similar past cases, reusing their solution and continuously retaining new cases into the case bank. There are some recent works \cite{DS-Agent, cbr-llm1,cbr-llm2} that integrate LLMs with CBR to enhance the capabilities of LLMs. Unlike these works, we explore the optimization of the CBR system in this work by further finetuning the retriever model and the LLM to align with the production scenario.
Notably, the Retrieve-Reuse process in the CBR system is similar to the well-established RAG techniques \cite{rag-1, rag-2, rag-survey}, offering valuable insights for optimizing the CBR system. However, prior approaches \cite{replug, dpa-rag} to finetuning the retriever model rely heavily on feedback from LLMs, which proves too costly for our production scenario. Moreover, previous RAG works \cite{raft, evidence-rag, simrag} focus on finetuning LLMs to generate responses with the retrieved document chunks in a robust manner, while the CBR system emphasizes adapting and reusing the retrieved cases for solving the new problem. As a result, these existing methods cannot be seamlessly leveraged for optimizing our CBR system.
\section{Conclusion}
In this paper, we explore the pioneering application of LLMs for functional test script generation. We introduce a CBR system that facilitates LLMs to effectively and flexibly utilize the mapping between test intent descriptions and function calls in the retrieved cases for test script generation. To further enhance the CBR system, we propose Re4, which consists of a reranking-based retrieval finetuning method for the retriever model and a reinforced reuse finetuning method for the LLM. Experimental results on two real-world datasets from Huawei Datacom demonstrate that the proposed CBR+Re4 approach significantly outperforms other baselines. Moreover, the Re4 method helps mitigate the issue of repetitive generation in LLMs, further enhancing the user experience.
\section{Limitation and Future Work}
Firstly, we utilize bge-m3 with cosine similarity as the retriever in this work, which can be further improved by hybrid retrieval methods, multiple-stage retrieval methods, etc. Secondly, more comprehensive offline evaluation metrics may help us better benchmark the performance of the algorithms, such as function parameter accuracy, execution success rate, etc. In this work, we focus on the evaluation of function correctness, as this is the most important business metric based on our preliminary internal investigation. In contrast, execution success rate is a much more reliable evaluation metric; however, it is hard to calculate due to the complicated execution environment required by the target commercial software. Thus, there is still a long way to go towards fully automated functional testing, and this work aims to serve as a good starting point. Furthermore, the proposed CBR system can also be extended to broader decision-making scenarios in fields such as law, medicine, and finance. We plan to further explore the potentials in these domains in future work. Lastly, this paper focuses on function correctness and does not perform comprehensive user-centric evaluation, such as user acceptance rate, user satisfaction rate, time savings, etc. Investigation on these metrics may help us understand the bottleneck of the algorithms better.

\begin{acks}
We truly thank the reviewers for their great effort in our submission. This work is supported by National Key R\&D Program of China under Grant (2023YFF0905400), National Natural Science Foundation of China through grants (624B2059, U2341229, 61976102, U19A2065, 62476110), Key R\&D Project of Jilin Province (20240304200SF), and International Cooperation Project of Jilin Province (20220402009GH).
\end{acks}

\bibliographystyle{ACM-Reference-Format}
\bibliography{references}

\appendix
\begin{figure*}[t]
    \includegraphics[width=\textwidth]{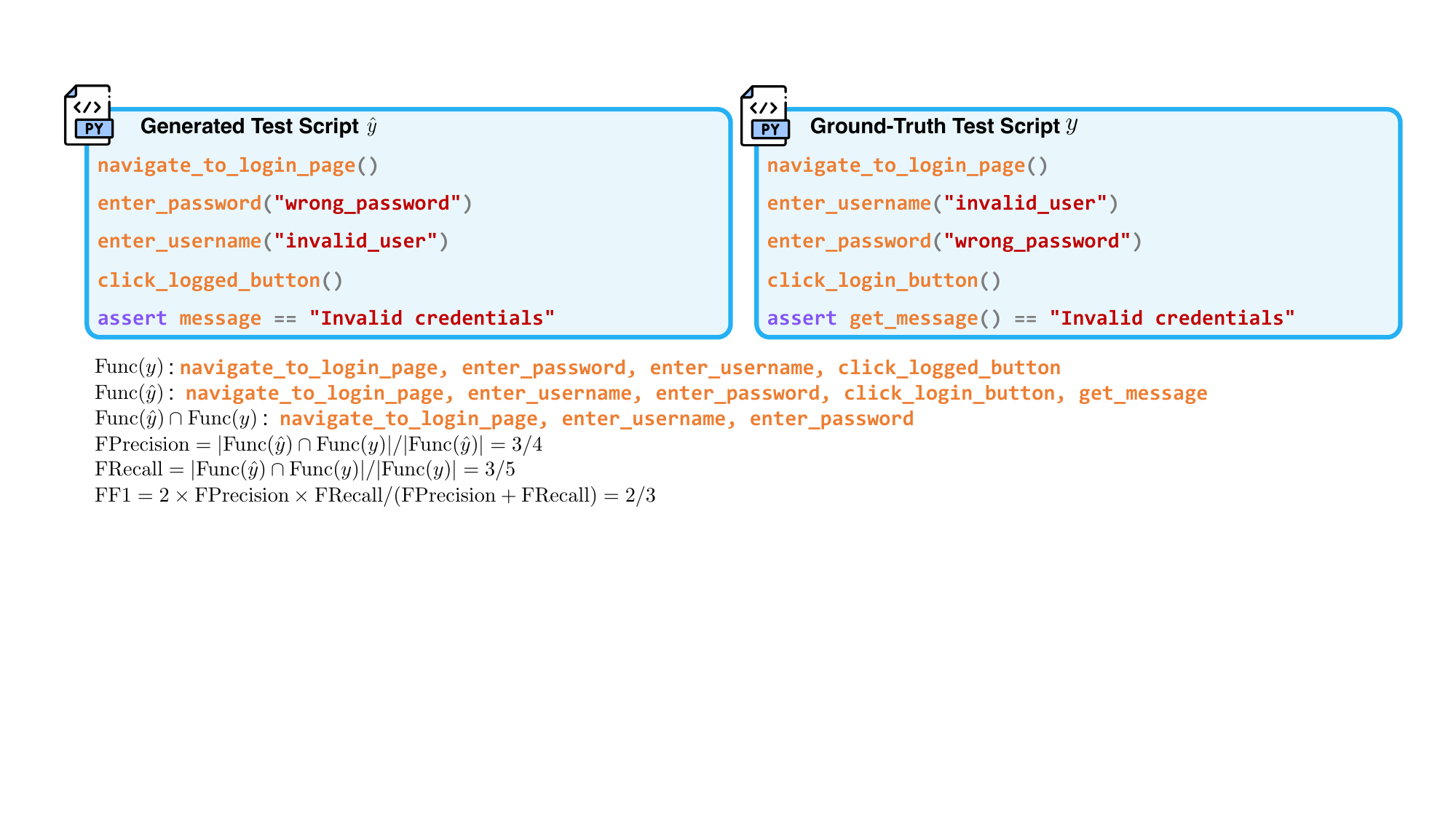}
    \caption{An example for the calculation of function precision, function recall and function f1 score.}
    \label{fig:metric-example}
    \vspace{-10pt}
\end{figure*}
\section*{Appendix}
\section{Algorithmic Details}
\subsection{Business Metric Details}
\label{app:metric}
For the proposed function precision, function recall, and function f1 score, we first extract the list of the invoked functions in the test scripts with abstract syntax tree, and then calculate these metrics with Eq. (\ref{eq:precision}), Eq. (\ref{eq:recall}), and Eq. (\ref{eq:f1}). To make it more clear, we provide an example of the calculation of these three metrics in Figure~\ref{fig:metric-example}.

Now, we detail the definition of the code similarity \cite{code-similarity} with the tool of Levenshtein distance \cite{levenshtein-distance}. Given the generated test script $\hat{y}$ and the ground-truth test script $y$, the code similarity is defined as:
\begin{equation}
    \text{CodeSimilarity}(\hat{y}, y)= 1 - \frac{d_{\text{L}}(\hat{y}, y)}{\max(|\hat{y}|, |y|)},
\end{equation}
where $d_{\text{L}}$ denotes the Levenshtein distance, measuring the minimum number of single-character edits (insertions, deletions, or substitutions) needed to change one string into another. As stated earlier, code similarity serves only as a complementary evaluation metric, and we take function f1 score as the most critical evaluation metric due to the characteristics of our production scenario.

\subsection{Algorithm Details}
\label{app:algorithm}
In this subsection, we provide the pseudo-codes for our proposed method. We first summarize the workflow of the proposed CBR system during deployment in Algorithm~\ref{alg:cbr-inference}. Our CBR system adopts the 4R cycle with Retrieve, Reuse, Revise and Retain. Then, we summarize the proposed reranking-based retrieval finetuning method in Algorithm~\ref{alg:retrieval}, which first generates pseudo labels for positive and negative examples and then finetunes the retriever model with contrastive learning. Finally, we summarize the proposed reinforced reuse finetuning method in Algorithm~\ref{alg:reuse}, which consists of supervised finetuning and reinforcement learning finetuning, striking the desired trade-off between efficiency and effectiveness.

\begin{algorithm}[h]
\caption{The CBR System During Deployment}
\label{alg:cbr-inference}
\begin{algorithmic}[1]
\STATE \textbf{Input:} Case Bank $\mathcal{C}$, Finetuned embedding model $\mathbf{E}_\phi$, and Finetuned LLM $\pi_\theta$.

\FOR{user request with test intent description $q$}
\STATE $\triangleright$ \textcolor{blue}{Retrieve}
\STATE Retrieve top-$M$ cases $\mathcal{C}_q$ based on Eq. (\ref{eq:retrieve})
\STATE $\triangleright$ \textcolor{blue}{Reuse}

\STATE Sample a test script $\hat{y} \sim \pi_\theta(\cdot|q,\mathcal{C}_q)$
\STATE $\triangleright$ \textcolor{blue}{Revise}
\STATE Revise the generated test script $\hat{y}$ as $y$ by test engineers if necessary
\STATE $\triangleright$ \textcolor{blue}{Retain}
\STATE Retain the test intent description and the test script as a new case into the case bank, i.e., $\mathcal{C}\leftarrow\mathcal{C}\cup \{(q,y)\}$
\ENDFOR
\end{algorithmic}
\end{algorithm}

\begin{algorithm}[h]
\caption{Reranking-based Retrieval Finetuning Method}
\label{alg:retrieval}
\begin{algorithmic}[1]
\STATE \textbf{Input:} Training set $\mathcal{D}$, Pretrained embedding model $\mathbf{E}_\phi$.
\STATE Initialize labeled training dataset $\mathcal{D}_{\text{labeled}}=\{\}$
\newline $\triangleright$ \textcolor{blue}{Generate pseudo-labels for positive and negative examples} 
\FOR{sample $(q,y)$ in $\mathcal{D}$}
\STATE Construct the case bank as $\mathcal{C}=\mathcal{D}\setminus\{(q,y)\}$
\STATE Retrieve top-$k$ cases $\mathcal{C}_q$ based on Eq. (\ref{eq:retrieve})
\STATE Rerank the retrieved cases $\mathcal{C}_q$ to generate pseudo-label for positive example $c^+$ and negative examples $\mathcal{C}^-$ as in Eq. (\ref{eq:pos}) and Eq. (\ref{eq:neg})
\STATE Store the labeled samples $\mathcal{D}_{\text{labeled}} \leftarrow \mathcal{D}_{\text{labeled}} \cup \{(q,c^+,\mathcal{C}^-)\}$
\ENDFOR
\newline $\triangleright$ \textcolor{blue}{Finetune the embedding model with labeled dataset}
\FOR{labeled sample $(q,c^+,\mathcal{C}^-)$ in $\mathcal{D}_{\text{labeled}}$}
\STATE Update $\phi$ by minimizing $\mathcal{L}(\phi)$ in Eq. (\ref{eq:infonce})
\ENDFOR
\end{algorithmic}
\end{algorithm}

\begin{algorithm}[h]
\caption{Reinforced Reuse Finetuning Method}
\label{alg:reuse}
\begin{algorithmic}[1]
\STATE \textbf{Input:} Training set $\mathcal{D}$, Finetuned embedding model $\mathbf{E}_\phi$, Large language model $\pi_\theta$.
\STATE Retrieve $M$ cases $\mathcal{C}_q$ for each sample in $\mathcal{D}$ using $\mathbf{E}_\phi$
\newline $\triangleright$ \textcolor{blue}{Supervised finetuning} 
\FOR{sample $(q,\mathcal{C}_q, y)$ in $\mathcal{D}$}
\STATE Update $\theta$ by minimizing $\mathcal{L}_{\text{SFT}}(\theta)$ in Eq. (\ref{eq:sft})
\ENDFOR
\newline $\triangleright$ \textcolor{blue}{Reinforcement learning finetuning}
\FOR{sample $(q,\mathcal{C}_q, y)$ in $\mathcal{D}$}
\STATE Sample a test script $\hat{y} \sim \pi_\theta(\cdot|q,\mathcal{C}_q)$
\STATE Update $\theta$ by minimizing $\mathcal{L}_{\text{REINFORCE}}(\theta)$ in Eq. (\ref{eq:reinforce})
\ENDFOR
\end{algorithmic}
\end{algorithm}

\section{Why Does REINFORCE Work in Our RLFT Setting?}
\label{app:reinforce-works}
REINFORCE \cite{reinforce} is the most basic on-policy RL algorithm. Despite its simplicity, it suffers from high variance in the stochastic gradient \cite{rloo, remax}. As highlighted in \cite{remax}, this high variance arises from two sources: (1) external randomness in the Markov Decision Process (MDP) and (2) internal randomness in the sampling of LLMs. In RLHF and RLFT settings, external randomness is eliminated due to the deterministic nature of the transition and reward functions in the MDP. However, in RLHF, external randomness still affects performance, as the model receives varying reward scales due to the diverse prompts in the training set, leading to high variance in the stochastic gradient \cite{remax}. In contrast, our RLFT setting does not exhibit this issue, since it focuses on enabling LLMs to reuse retrieved cases for test script generation, significantly narrowing the action space. This results in more aligned reward scales and reduced variance.

\begin{figure}
    \includegraphics[width=0.5\textwidth]{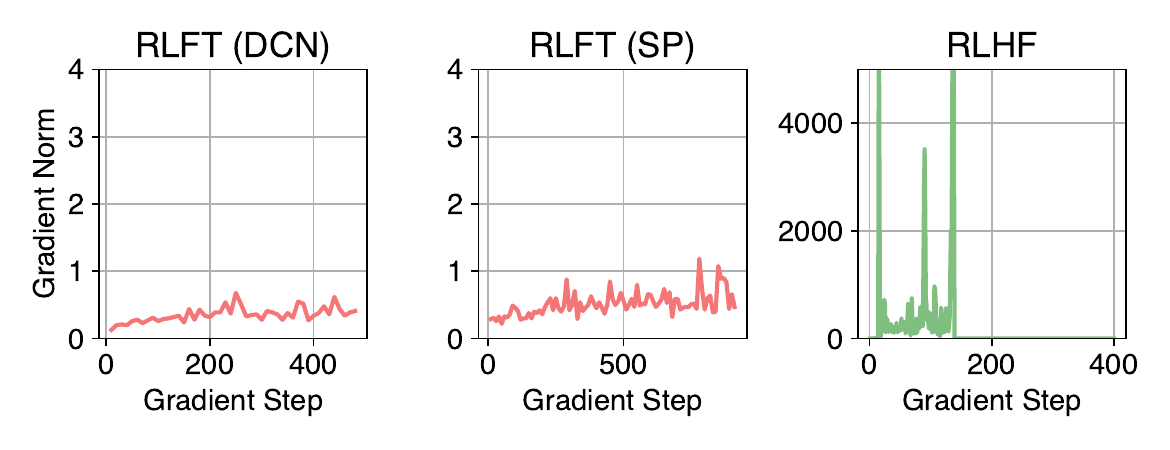}
    \caption{Comparison for the gradient norm of REINFORCE in RLFT and RLHF.}
    \label{fig:grad-norm}
\end{figure}

Now, we provide empirical evidence supporting the above claim. Since a smaller gradient variance corresponds to a smaller gradient norm, we follow prior work \cite{remax} and plot the gradient norm of REINFORCE in our RLFT setting and RLHF setting in Figure~\ref{fig:grad-norm}. For RLHF, we follow \cite{remax} to train the LLM with the consistent training setups as ours, where we adopt three prompt dataset (\textit{ultrafeedback}, \textit{lmsys-chat-1m}, \textit{sharegpt-en}) and the reward model (UltraRM-13B) for RLHF. The results reveal that the gradient norm of REINFORCE in our RLFT settings remains below 1 across both datasets, whereas this value fluctuates between 10 and 4000 in RLHF. Thus, REINFORCE does not suffer from the high variance issue and can achieve the desired performance in our RLFT setting. Note that the gradient norm of REINFORCE in RLHF remains 0 in the later stage of training due to training collapse. 

\section{Discussion of the Compared RLHF Algorithms}
\label{app:rlhf}
In this section, we present a detailed discussion on how we tailor the state-of-the-art RLHF algorithms compared in the main body of the paper for our RLFT setting. 

\noindent\textbf{Online DPO \cite{online-dpo}} samples two test scripts $\hat{y}_1$ and $\hat{y}_2$ with the LLM in an on-policy manner. Then, we can label the preference $y_w \succ y_l$ with the script similarity such that $\text{FF1}(y_w,y)>\text{FF1}(y_l,y)$. The optimization process aligns with the standard DPO loss \cite{dpo} as:

\begin{equation}
    \mathcal{L}_{\text{DPO}}(\theta)=-\log\sigma\left(\beta\log\frac{\pi_\theta(y_w|q,\mathcal{C}_q)}{\pi_{\theta_{\text{SFT}}}(y_w|q,\mathcal{C}_q)}-\beta\log\frac{\pi_\theta(y_l|q,\mathcal{C}_q)}{\pi_{\theta_{\text{SFT}}}(y_l|q,\mathcal{C}_q)}\right).
\end{equation}
However, Online DPO only reserves the preference relationship for alignment and ignores the fine-grained golden reward information, thereby decreasing the sample efficiency and resulting in inferior alignment performance. Moreover, it requires two on-policy samples per query, which brings $2\times$ time and computational costs compared to REINFORCE.

\noindent\textbf{Remax \cite{remax}} samples two test scripts $\hat{y}_1$ and $\hat{y}_2$ with the LLM in an on-policy manner. Among them, $\hat{y}_2$ is sampled via greedy decoding to serve as a baseline for variance reduction. The optimization process merely utilizes $\hat{y}_1$, with the standard REINFORCE \cite{reinforce} loss function as:

\begin{equation}
    \mathcal{L}_{\text{Remax}}(\theta)=-\log\pi_\theta(\hat{y}_1|q,\mathcal{C}_q)[r(\hat{y}_1)-r(\hat{y}_2)].
\end{equation}
However, the introduction of baseline value $r(\hat{y}_2)$ can be a redundant design in our context, and may even lead to a biased gradient estimator, despite being theoretically unbiased in expectation. Different from open-ended text generation, our setting only requires LLMs to reuse the retrieved cases for test script generation, thus significantly narrowing the action space. As such, the internal randomness mentioned in \cite{remax} is much less problematic in our setting. Similar to Online DPO, Remax also requires two on-policy samples per query, which brings $2\times$ time and computational costs compared to REINFORCE.

\noindent\textbf{RLOO \cite{rloo}} samples $K$ test scripts per query with the LLM in an on-policy manner. The optimization process utilizes the standard REINFORCE \cite{reinforce} with leave-one-out estimator for variance reduction, which can be formulated as:
\begin{equation}
    \mathcal{L}_{\text{RLOO}}(\theta)=-\frac{1}{K}\sum_{i=1}^{K}\left[\log\pi_\theta\left(\hat{y}_i|q,\mathcal{C}_q\right)\left(r\left(\hat{y}_i\right)-\frac{1}{K-1}\sum_{j \neq i}r\left(\hat{y}_k\right)\right)\right].
\end{equation}
Similarly, the variance reduction technique is also redundant for our setting. As suggested in \cite{rloo}, we set $K=4$ in our implementation, which brings $4\times$ time and computational costs compared to REINFORCE.

\noindent\textbf{GRPO \cite{grpo}} also samples $K$ test scripts per query. In contrast to RLOO, GRPO adopts the loss function similar to PPO \cite{ppo} for finetuning, which can be formulated as:
\begin{align}
    &\mathcal{L}_{\text{GRPO}}(\theta)=-\frac{1}{K}\sum_{i=1}^{K}\frac{1}{T_i}\sum_{t=1}^{T_i}\nonumber\\ &\left[
    \min\left(\hat\rho_{i,t}
    \hat{A}_{i,t}, \text{clip}(\hat\rho_{i,t}
    \hat{A}_{i,t}, 1-\epsilon,1+\epsilon)\right)-\beta\mathbb{D}_{\text{KL}}(\pi_\theta||\pi_{\theta_{\text{SFT}}})\right],
\end{align}
where $T_i$ denotes the number of tokens in $i$-th generated script, $\hat\rho_{i,t}=\frac{\pi_\theta(\hat{y}_{i,t}|q,\mathcal{C}_q,\hat{y}_{i,<t})}{\pi_{\theta_{\text{old}}}(\hat{y}_{i,t}|q,\mathcal{C}_q,\hat{y}_{i,<t})}$ denotes the importance rate of the $t$-th token in $i$-th generated test script, $\hat A_{i,t}= \frac{r(\hat{y}_i)-\text{mean}(\{r(\hat{y}_1),...,r(\hat{y}_K)\})}{\text{std}(\{r(\hat{y}_1),...,r(\hat{y}_K)\})}$ denotes the group-relative reward for advantage estimation, $\epsilon$ denotes the clipping parameter. GRPO achieves better performance than the proposed REINFORCE algorithm, benefiting from $K\times$ on-policy samples per query. Different from \cite{grpo} that sets $K=64$, we set $K=4$ in our implementation to accommodate computational constraints, which brings $4\times$ time and computational costs compared to REINFORCE.

\begin{figure}
    \includegraphics[width=0.45\textwidth]{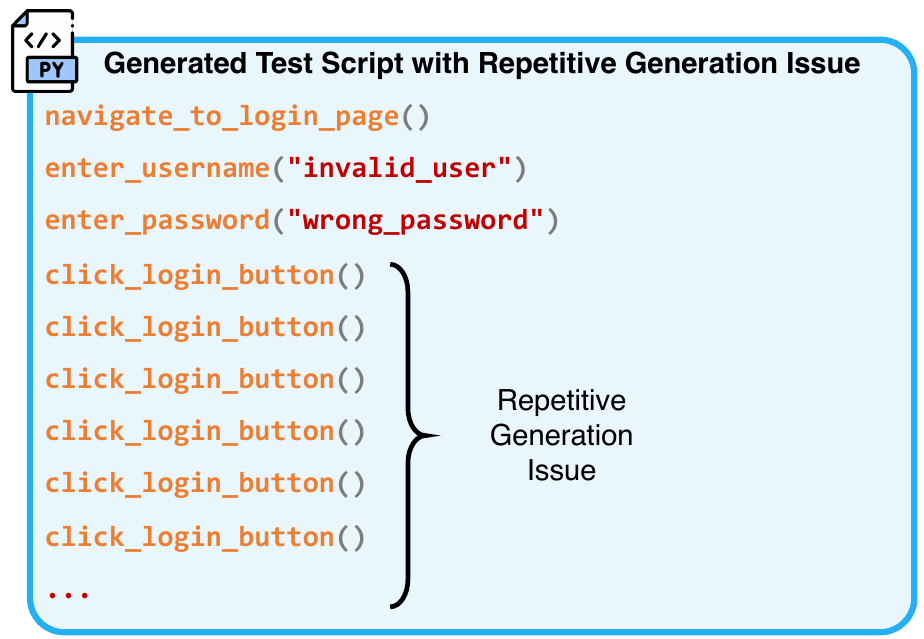}
    \caption{An example of generated test script with repetitive generation pattern.}
    \label{fig:repetitive-example}
\end{figure}

\section{Example of Repetitive Generation Issue}
\label{app:rgi}
Due to business considerations, we are unable to present a realistic example from our production scenario. We construct an illustrative example, as shown in Figure \ref{fig:repetitive-example}. Here are some possible reasons for this phenomenon: (1) Poor coding practices by human test engineers result in ground-truth test scripts with repetitive patterns, which is memorized by the LLMs through the SFT objective. (2) Noise within the SFT objective, as illustrated in Figure~\ref{fig:motivation}, contributes to hallucination issues, potentially worsening the repetitive generation problem.

\end{document}